\documentclass[12pt,preprint,rotate]{aastex}

\shorttitle{Kinematics of Stars in Kapteyn Selected Area 71}
\shortauthors{Casetti-Dinescu, et al.}

\begin{document}

\title{Kinematics of Stars in Kapteyn Selected Area 71: Sampling the Monoceros and Sagittarius
Tidal Streams}

\author{Dana I. Casetti-Dinescu\altaffilmark{1,2,3}, Jeffrey L. Carlin\altaffilmark{4}, 
Terrence M. Girard\altaffilmark{2}, Steven R. Majewski\altaffilmark{4}, Jorge Pe\~{n}arrubia\altaffilmark{5} and
Richard J. Patterson\altaffilmark{4}}

\altaffiltext{1}{Astronomy Department, Van Vleck Observatory, Wesleyan University, Middletown, CT 06459}
\altaffiltext{2}{Astronomy Department, Yale University, P.O. Box 208101,
New Haven, CT 06520-8101}
\altaffiltext{3}{Astronomical Institute of the Romanian Academy, Str.
Cutitul de Argint 5, RO-75212, Bucharest 28, Romania}
\altaffiltext{4}{Department of Astronomy, University of Virginia,  P.O Box 400325, Charlottesville, VA 22904-4325}
\altaffiltext{5}{Department of Astronomy, University of Victoria, Victoria, BC, Canada}

\begin{abstract}
We present a 3D kinematical analysis of stars located in Kapteyn Selected Area 71 ($l = 167.1\arcdeg, b = -34.7\arcdeg$),
where previously a stellar excess was found (Dinescu et al. 2002, Newberg et al. 2002). Previous findings indicated that
the stellar excess has a cold kinematical signature as inferred from proper motions, and was initially 
associated with debris from the Sagittarius dwarf galaxy (Sgr) --- namely the southern trailing tail. 
We have obtained radial velocities using the Hydra multiobject spectrograph on the WIYN 3.5 m telescope. Results 
for 183 proper-motion selected stars indicate that the dominant population in this 
stellar excess is not debris from Sgr, but rather a population that kinematically belongs to the 
ring-like stream that is now known as the Monoceros stream. The orbit determined for this population agrees 
very well with the predictions for the Monoceros stream from Pe\~{n}arrubia et al. (2005). The radial-velocity
dispersion of this population is between 20 and 30 km/s, lower than that of the Galactic field. Also,
the shape of the radial-velocity distribution shows a sharp cut-off on one side, which is more
in line with model predictions of the disruption of a satellite rather than with the distribution of the Galactic
field. Despite the fact that we now believe most of the stars in the stellar excess to be part of Monoceros,
about ten stars in this stellar excess have highly negative radial velocities, which is a clear indication
of their membership to the Sgr trailing tidal tail. 

\end{abstract}


\section{Introduction}

The wealth of streams and overdensities discovered recently in the Milky Way halo
(e.g., Belokurov et al. 2006, Grillmair 2006, Vivas and Zinn 2006) while spectacular,
 has made our task of understanding the halo more challenging than some ten years ago.
A thorough knowledge of these structures' kinematics is key to understanding their origin and
significance as building blocks of the Galaxy in the hierarchical view of its formation.

In this contribution, we focus on a region in the sky where two major tidal streams
appear to reside at different distances: the Sagittarius (Sgr) tidal stream, 
and the Monoceros stream/structure (hereafter Mon).
While the Sgr tidal stream is better understood, Mon has a more ambiguous origin because 
its progenitor has not yet been established. Mon was ``pieced together" by the discovery of 
stellar overdensities roughly parallel to the Galactic plane, and above and below the plane.
Many large-area photometric surveys have contributed to the mapping of Mon: 
starting with SDSS (Newberg et al 2002, Yanny et al. 2003) and continuing with 2MASS (Rocha-Pinto et 
al. 2003), the INT/WFC (Ibata et al 2003, Conn et al. 2005) and the AAT/WFI (Conn et al. 2007). 
It was realized by Rocha-Pinto et al. (2003) and Ibata et al. (2003), that this structure is immense,
spanning from $l=100\arcdeg$ to $270\arcdeg$, and that it is located in a ring-like form at a
Galactocentric distance of $\sim 18$ kpc. 

Our region of interest is a $40\arcmin\times40\arcmin$ area centered on
Kapteyn Selected Area (SA) 71,
at $(l,b) = (167.1\arcdeg, -34.7\arcdeg)$.
The location of SA 71 is within the general area
of the Monoceros stream detections (see e.g., Fig. 3 in Rocha-Pinto et al. 2003).
Our previous study of this region (Dinescu et al. 2002, hereafter D02) 
included the determination and analysis of
absolute proper motions and
photographic photometry of stars within $17 < V < 20$.  D02 have demonstrated that SA 71 contains an
excess of blue stars when compared to a Galactic model that was well fit in three other
lines of sight. The proper motions of the excess stars show a tight clump, also supporting
the notion of a kinematically cold substructure. 
At that time, the only well-known tidal 
stream was that of Sgr, and D02 have interpreted their data as the Sgr stream. Specifically, the 
trailing tidal tail that arches underneath the Sun's location to reach the Galactic plane 
toward the Anticenter direction (e.g., Majewski et al. 2003) best explained the data in D02:
Galactic location, colors and magnitudes of excess stars, and approximately the mean absolute proper motion.
Here, we present radial velocities of stars selected in the excess region of the
color-magnitude diagram (CMD) and in the proper-motion diagram (also known as the 
vector-point diagram, VPD), 
and show that, while a small number of stars in this excess belong to Sgr, most of them are part
of a different structure. The kinematical properties of this structure are now best explained
in the general context of the Monoceros structure. 

The current study is part of a larger proper-motion and radial-velocity survey in over
50 SAs. The characteristics of this survey, as well as the astrometric techniques
are throughly described in Casetti-Dinescu et al. (2006, CD06).
We will therefore refer the interested reader to that contribution for more specific aspects 
of the reductions. All proper motion units throughout the paper are mas/yr, and radial velocities are
heliocentric, unless otherwise specified.
 
The paper is structured as follows. In Section 2 we present our data: photometry, proper motions
and radial velocities, in Section 3 we discuss our results and 
compare the observed kinematics with the predictions of 
the Besan\c{c}on Galactic model (Robin et al. 2003). In Section 4 we compare our results
with the predictions of the
Pe\~{n}arrubia et al. (2005, hereafter P05) model of the Mon as a disrupting satellite.
In Section 5 we discuss the possibility that the Galactic warp and/or other nearby overdensities may be 
present in SA 71. Section 6 outlines our conclusions. 

\section{The Data}

\subsection{Photometry and Proper Motions}

The set of 14 photographic plates used in this study
includes 11 Mayall 4 m plates (epoch=1975 and 1996), 
two Du Pont 2.5 m plates (epoch=1996), and one Mt. Wilson 60'' plate (epoch=1912).
The Du Pont and Mt. Wilson plates contain two exposures each: a 3 minute and a 60 minute.
All plates were scanned with the Yale PDS microdensitometer, based on an input list of objects.
This list was determined by combining coarser scans of six Mayall plates, which are the deepest
in our set. Both the photographic photometry and the proper-motion determinations were 
described in D02 and CD06, and therefore we refer the reader to these papers for
detailed measurements and reduction procedures. The photographic photometry was calibrated to
standard Johnson $BV$ magnitudes using CCD photometry obtained with the Swope 1 m telescope
at Las Campanas in 2001 (see D02). The CCD photometry covered $23\%$ of the total area 
covered by the photographic plates and to a limiting magnitude $B = 22$, while the Mayall 4 m photographic
plates are deeper by 0.7 mags in $B$. From the comparison of photographic
and CCD magnitudes, and assuming that the CCD magnitude errors are substantially
smaller than the photographic ones, we have estimated that the photographic $V$ uncertainties are 
0.05 mags for $V = 17$ to 21, and the $B-V$ uncertainties are 0.08 mags, for the same 
$V$ range. Outside this $V$-magnitude range, the uncertainties increase rapidly due to saturation
of the photographic images at the bright end, and due to the decrease in $S/N$ at the faint end.

In this work, we have relied on the SExtractor resolution classifier (Bertin \& Arnouts 1996) for
object classification, using the deepest, best-seeing, Mayall 4 m, modern-epoch plate ($\#4323$,  IIIa-F + GG495).  
The value of the stellarity index that separates galaxies from stars was determined from 
a number of criteria: 1) visual
inspection of images, 2) plots of image parameters such as image radius and
peak density, and 3) comparison with the 2MASS classifier for objects with $V < 19$.
For another SA field ---  namely SA 94 --- that has similar plate material to
 SA 71  (see CD06), together with SDSS photometry and 
object classification, we can also compare our object classification with that of SDSS.
We obtain that our classification of galaxies recovers $75\%$ of the SDSS galaxies at $V = 21.0$ and
$50\%$ at $V = 21.6$. 

The astrometric reductions have been redone in this work --- compared to that of D02 --- to
include the measurements of the old Kapteyn, Mt. Wilson plate. This expands the time baseline from
$\sim 20$ to 84 years, but only for stars brighter than $V \sim 18$.
Also, we have used a list of well-measured,
round-shaped galaxies to model magnitude-dependent systematics in the proper motions. 
The reduction procedure has been described in D02 and in Sections 2.5 and 2.6 of CD06, and will not be
repeated here. The galaxies' proper motions (or positional residuals) do not show significant
trends as a function of color, for $0.3 < B-V < 1.8$, or magnitude, in the range $V = 18$ to 21.5.
The correction to absolute proper motion is determined by galaxies, selected using the following criteria:
$V= 18$ to 21.5, measured on at least six plates, and with proper-motion size less than 30 mas/yr.
This sample includes 618 galaxies.
We have then used probability plots (Hamaker 1978) trimmed at $10\%$ on this sample to estimate its mean and dispersion.
We obtain an average relative proper motion for the galaxy sample of 
$(\mu_{\alpha}~cos~\delta, \mu_{\delta}) = (-2.32 \pm 0.18, 2.28 \pm 0.19)$ mas/yr, which 
is applied as an offset to all objects; this number is also known as the correction to absolute proper motion.

For stars, we obtain formal proper-motion uncertainties of 0.5 mas/yr for $V = 18$, 0.9 mas/yr for 
$V = 19$, 1.4 mas/yr for $V = 20$, and 2.2 mas/yr for $V = 21$, in each coordinate.

\subsection{Radial Velocities}

\subsubsection{Sample Selection}

The stars selected for followup spectroscopy are within the magnitude and color
ranges where the stellar excess was found in D02. Specifically,
the approximate color range is $0 < (B-V) < 1.1$ and magnitude range is $18 < V < 19.5$. Likewise,
stars were selected within a $\sim 3$ mas/yr radius from the proper-motion center of the
tight clump defined by the excess, blue stars. A few stars outside these ranges were also
considered, for comparison purposes. The program stars for spectroscopy are
shown in Figure 1 as bold symbols among the general stellar population of the
field; selections in the color-magnitude diagram (CMD) and the vector-point diagram (VPD) are
shown.

\begin{figure}[tbh]
\includegraphics[scale=0.70,angle=-90]{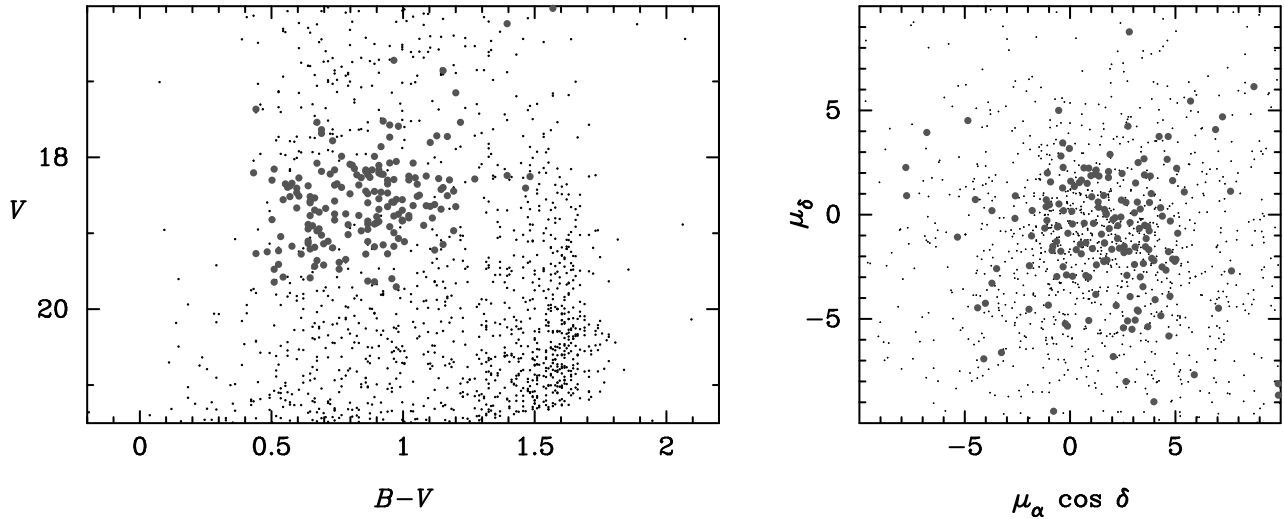}
\caption{The selection of stars (filled circles) for RV measurements in the CMD (left) and in the
VPD (right).}
\end{figure}

\subsubsection{Observations}

Radial velocities (RVs) were obtained from observations taken with the 3.5 m WIYN 
telescope during six observing runs. We have used the 
Hydra multifiber positioner with the WIYN bench spectrograph in two
setups. The first one (Dec. 2002, Nov. 2003 observing runs) used the 800@30.9 grating with 
the red fiber cables and an order centered in the 
neighborhood of the Mg $b$ triplet (5170 \AA), and covering about 980 \AA$~$
of the spectrum.
This corresponds to a dispersion of 0.478 \AA$~$pix$^{-1}$, and a 
%resolving power $R \sim 5400$. 
spectral resolution of 0.956 \AA.
The comparison source was a Cu-Ar lamp.
The second spectrograph configuration (Dec. 2005, Oct. 2006,
Dec. 2006 and Dec. 2007) used the 600@10.1 grating with the red fiber cable to yield
a wavelength coverage $\lambda$ = 4400--7200 \AA~at a dispersion of
1.397 \AA$~$pix$^{-1}$, for a spectral resolution of 3.35 \AA.  
Each of these 2005-7 datasets was obtained in less than optimal conditions,
including substantial scattered moonlight in Dec. 2005, and cloudy
conditions in the 2006 and 2007 runs.

Table 1 contains the summary of the observations as follows:
the first column lists the dates of the observing runs, the second column the number of 
Hydra configurations observed during the respective run, the third column the total exposure time, the fourth
column the central wavelength, the fifth column the dispersion. The sixth column
indicates the number of RV standard stars observed, and the total number of individual standard star spectra taken.
The seventh column lists the number of stars with a Tonry-Davis ratio (TDR, Tonry \& Davis 1979) 
larger or equal to 8, and the last column shows the $V$-magnitude range of the program stars observed in each run.
Each Hydra configuration was exposed in sets of 30 min, to obtain the total exposure
time listed.
For RV calibrations, each night we observed standard stars which were typically 
RV-calibrated HD stars listed in the Astronomical Almanac as well as stars 
in M 67 (Mathieu et al. 1986) selected to be single, well-measured stars.
%JLC added:
On the 2005-6 observing runs, each individual HD standard was observed
through multiple Hydra fibers, to yield multiple, distinct cross-correlation
template spectra (with that total number also listed in Table 1). The magnitude range of
the standard stars is $V = 7.3$ to 11.6, therefore these are well-measured stars with 
high S/N.
%
%SRM: Should we say something about these being very HIGH S/N, like S/N > 70, or whatever?

\begin{table}[htb]
\begin{center}
\caption{Summary of Observations}
\begin{footnotesize}
\begin{tabular}{cccccccc}
\tableline
\\
\multicolumn{1}{c}{Date} & \multicolumn{1}{c}{Config.} & \multicolumn{1}{c}{Exp.} & \multicolumn{1}{c}{Wavelength} & 
\multicolumn{1}{c}{Dispersion} & \multicolumn{1}{c}{RV standards} & \multicolumn{1}{c}{N$_{TDR \ge 8}$} & 
\multicolumn{1}{c}{$V$-range} \\
 & & (hours) & (\AA) & ( \AA$~$pix$^{-1}$) & (\# of distinct spectra) & & \\
\tableline
\\
Dec 2002 &    1   &    2    &  5170   & 0.478 & 2 HD (2) & 37 & 18-19 \\
Nov 2003 &    2   &    5, 2 &  5170   & 0.478 & 2 HD+15 M67 (17) & 74 & 18-19 \\
Dec 2005 &    1   &    3    &  5800   & 1.397 & 1 HD (5) & 27 & 18.5-19.5 \\
Oct 2006 &    1   &    3    &  5800   & 1.397 & 11 HD (65) & 47 & 18.2-19 \\
Dec 2006 &    1   &    2    &  5800   & 1.397 & 2  HD (10) & 58 & 16.7-19.3 \\ 
Dec 2007 &    1   &    4    &  5800   & 1.397 & 9 HD (40) & 55 & 16.0-19.7 \\ 
\tableline
\end{tabular}
\end{footnotesize}
\end{center}
\end{table}

%JLC: I added the (total # spectra) thing in the table above, but
%rather than having this in the header, should this table have a
%legend explaining everything?

The data were reduced using the DOHYDRA reduction package in IRAF
after the standard pre-processing was applied to the initial two-dimensional 
spectra. For the wavelength calibration we have used 15-17 comparison lines.  
The sky subtraction was done outside the DOHYDRA package,
for each observation individually, with typically 10 fibers to define 
it. Then, for each Hydra configuration, the multiple sets of 30-minute
 observations were median combined. Finally, RVs were derived 
using the FXCOR task in IRAF to cross-correlate these target spectra against all RV standard spectra.

\begin{figure}[!h]
\includegraphics[scale=0.80]{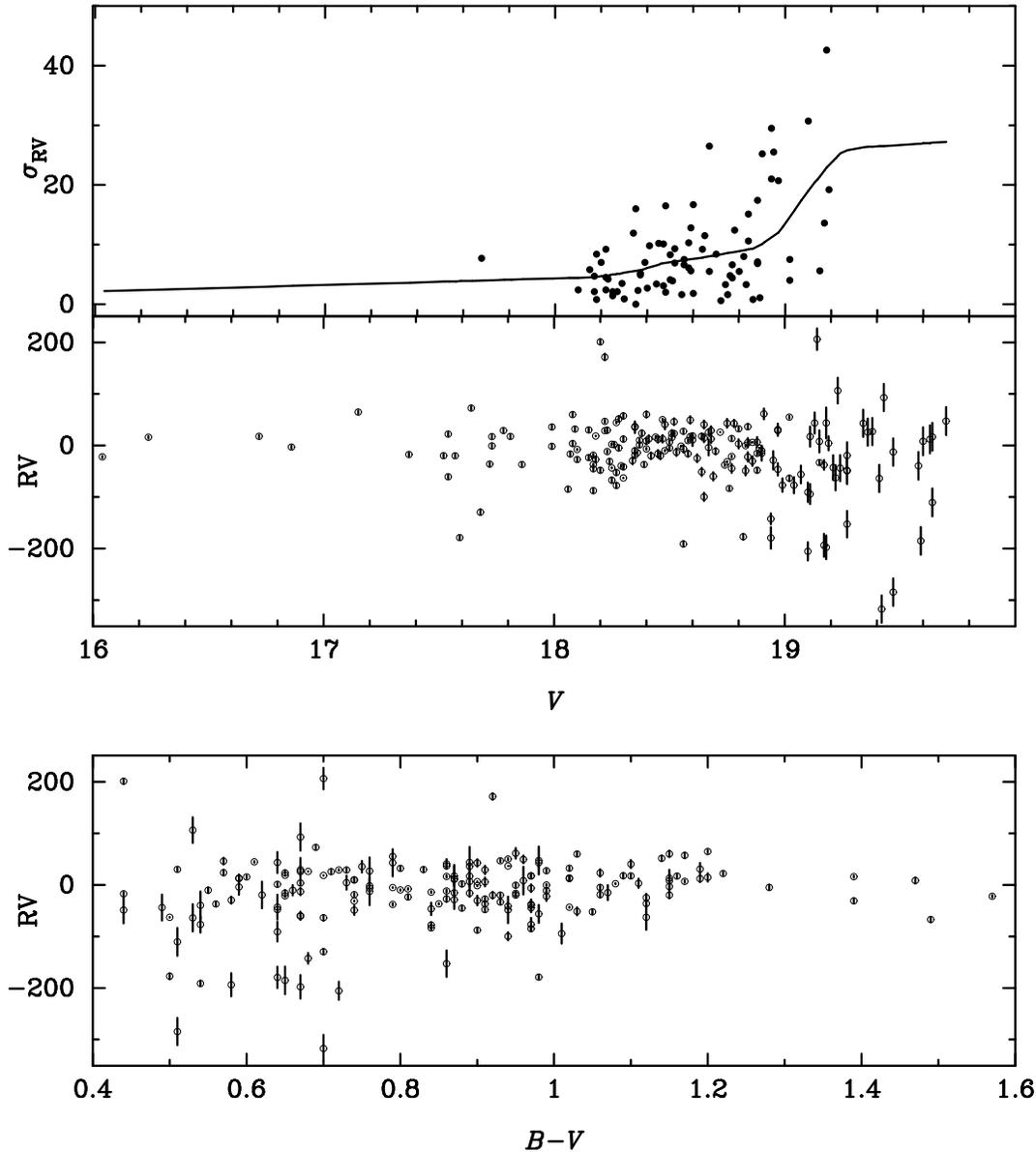}
\caption{The RV scatter for stars with multiple measurements as a function of magnitude (top).
The final RVs as a function of magnitude (middle) and as a function of color (bottom)
for all of the stars in our sample. Units are km/s.}
\end{figure}

Comparing the RVs obtained from our observations of the standard stars during the 2003 run  
with those from the literature, we obtain that 
the RV difference is $0.03 \pm 0.26$ km/s, 
and the standard deviation is 1.1 km/s. As mentioned above, these are bright
($V = 7.3$ to 11.6), well-measured stars.
Our program stars are however much fainter (see Tab. 1). Therefore, we have kept 
only those determinations with a TDR $\ge 8$ ($S/N \ge 15$).
%
%SRM: Is this S/N per pixel, or per resolution element (and same question throughout 
% the paper when S/N's are mentioned)?  But I find it hard to believe that you got useful
% RVs from S/N = 5 spectra, no matter how S/N is being defined.   Usually you need about 
% 100 counts/pixel before the spectra are reliable...
%

The Hydra configurations were designed to have some overlap between different
sets of observations. Comparing results from different runs, where the overlap 
includes between 7 and 20 stars, we obtain the following. The 
RV errors are between 5 and 10 km/s per star, per single measurement for $V = 18$ to 18.8.
For fainter stars, the errors increase rapidly to reach $\sim 20$ km/s at
$V \sim 19.0$. 
We have also found significant (2 to 5$\sigma$) RV offsets between different runs;
these are of the size of 5 to 10 km/s.
%JLC: Table 2 shows PMs - is there a table you need to add in?  
%Also, I don't like the statement 'are probably due to an insufficient
%calibration with standards', since we've presumably checked that our
%standards are well-measured.  I think simply mentioning systematic
%offsets without speculating about their origin is OK.
%
% SRM: I had the same reaction -- that it seemed this "insufficient calibration" was not a 
% great reason.  I wonder if there were differences in the wavelength calibration, or something
% like that that caused problems.  I presume that JLC is the one who changed this to "Tab. 1".
%
Our two best RV-standard calibrated runs, the 2003, 5-h integration 
set, and the Oct. 2006 set, have an offset of $ -1.2 \pm 3.2$ km/s 
for eight stars in common between the two runs.  To place all 
observations on the same system, we have chosen the 2003, 5-h integration run
as our ``reference'' system, and we have offset the other sets onto this system.
For stars with multiple RV determinations, we take the average as the final RV
%JLC Is this 'average' a weighted average?  Seems like it should be, if not.
value, and the RV error is calculated from the scatter of the determinations.
For stars with a single RV determination, we adopt RV errors as follows:
First, from stars with multiple determinations, we plot the RV scatter as a function of $V$
magnitude. We then apply a moving median (0.5 mag bin) to derive a curve that describes the dependence
of RV error as a function of magnitude. Finally, for each star of a given magnitude, we interpolate
the derived dependence of RV error with magnitude to determine the RV error.
In Figure 2 (top), we show the RV scatter as a function of magnitude for stars with multiple
determinations, and the adopted curve for stars with single determinations. The middle plot
shows the RVs with their error bars as a function of magnitude, while the bottom plot shows the
RVs as a function of $B-V$ colors. Our entire sample contains 183 stars.

%SRM: Presumably the RV precision is more a function of S/N than magnitude, and that 
% not all stars of a given magnitude (and color!) are observed to the same S/N?   

%SRM: WHy is there no table of the RV, proper motion, color and magnitude values??

\section{Discussion of the Observations}

We proceed now to interpret our results in light of the three most 
plausible contributors to the stellar counts
in this region of the sky: the Galactic field, the Sgr tidal stream 
and the Monoceros stream.

To compare our data with expectations for the stellar Galactic field, we have used the
simulated catalogs obtained with the Besan\c{c}on Galactic model (Robin et al. 2003).
The simulated data were generated in a 0.43 degree square field for stars
between $V = 14$ to 22. The photometry of the model was convolved with 
a 0.05 $V$-mag error and a 0.08 mag color error in $(B-V)$, which is a good representation
of our uncertainties between $V = 17$ to 21, where the comparison is made (see Section 2.1). 
The proper-motion errors to be convolved with the model are derived from the run of
observed proper-motion errors as a function of magnitude (Section 2.1). After
eliminating all stars with formal errors larger than 4 mas/yr, we
fit a second-order polynomial to the observed dependence of proper-motion errors with magnitude 
%
%SRM: Not 100% clear here -- "this observed dependence" of what on what?  You mean p.m. errors
% with magnitude?
%
in the range $V = 14$ to 22. The coefficients thus obtained are input into the Besan\c{c}on model.
Similarly, for radial velocities, we determine the quadratic dependence of formal
errors as a function of magnitude as given by the data in the range $V = 17.5$ to 19
(Section 2.2) and input this function into the Besan\c{c}on model.

For the Sgr stream, we use the disruption model within a
spherical potential from Law et al. (2005). In this contribution, we do not fully analyze
the kinematical data for Sgr, but rather use the Law et al. model as a guideline for identifying
the Sgr population and its likely kinematics. A future contribution will be entirely
dedicated to the kinematical analysis of the Sgr southern trailing tail, and it will include 
more SA fields besides SA 71.

For the Mon disruption model we use the predictions
from P05, and  dedicate Section 4 to this kinematical comparison.  

\subsection{What is Creating the SA 71 Starcount Excess ?}

In our D02 paper we have compared $(B-V)$ color distributions in three $V$-magnitude bins
with those predicted by another Galactic model (M\'{e}ndez \& van Altena 1996) in four
SA fields, including SA 71. We have shown there that, while three of the SA fields
have color distributions that agree with the Galactic model, SA 71 has excess
counts at blue colors, ($B-V) < 1.1$. Specifically, the excess counts were found to 
peak at $(B-V) = 1.0$ for $V = 18 - 19$,
 and at $(B-V) = 0.6$ for $V = 19 - 20$ 
(see Fig. 1 in D02). In the bright magnitude bin, the observations indicated 219 stars, and the 
model predicted 167, while in the faint magnitude bin, the observations found 256 stars, while the 
model predicted 210.

In this contribution, we will not repeat the starcount analysis from D02, as we are mainly interested in the 
kinematical aspect. Instead, we show in Figure 3
the CMDs as given by the Besan\c{c}on model (top) and by our data (middle) for a qualitative
comparison. We have applied a $V=22.0$ magnitude limit in our data, since the quality
is rather poor beyond this value. For the model, which was initially generated within
$V = 14$ to 22, we have applied a $B=22.9$ limit, to mimic
the blue photographic plate limit of the data.
%SRM: I'm confused, aren't both showing a B=22.9 limit?
From Figure 3, the excess of stars in the observations compared to the model is 
readily visible at $(B-V) < 1.1$ and $ V \ge 18$. 

\begin{figure}[!h]
\begin{center}
\includegraphics[scale=0.80]{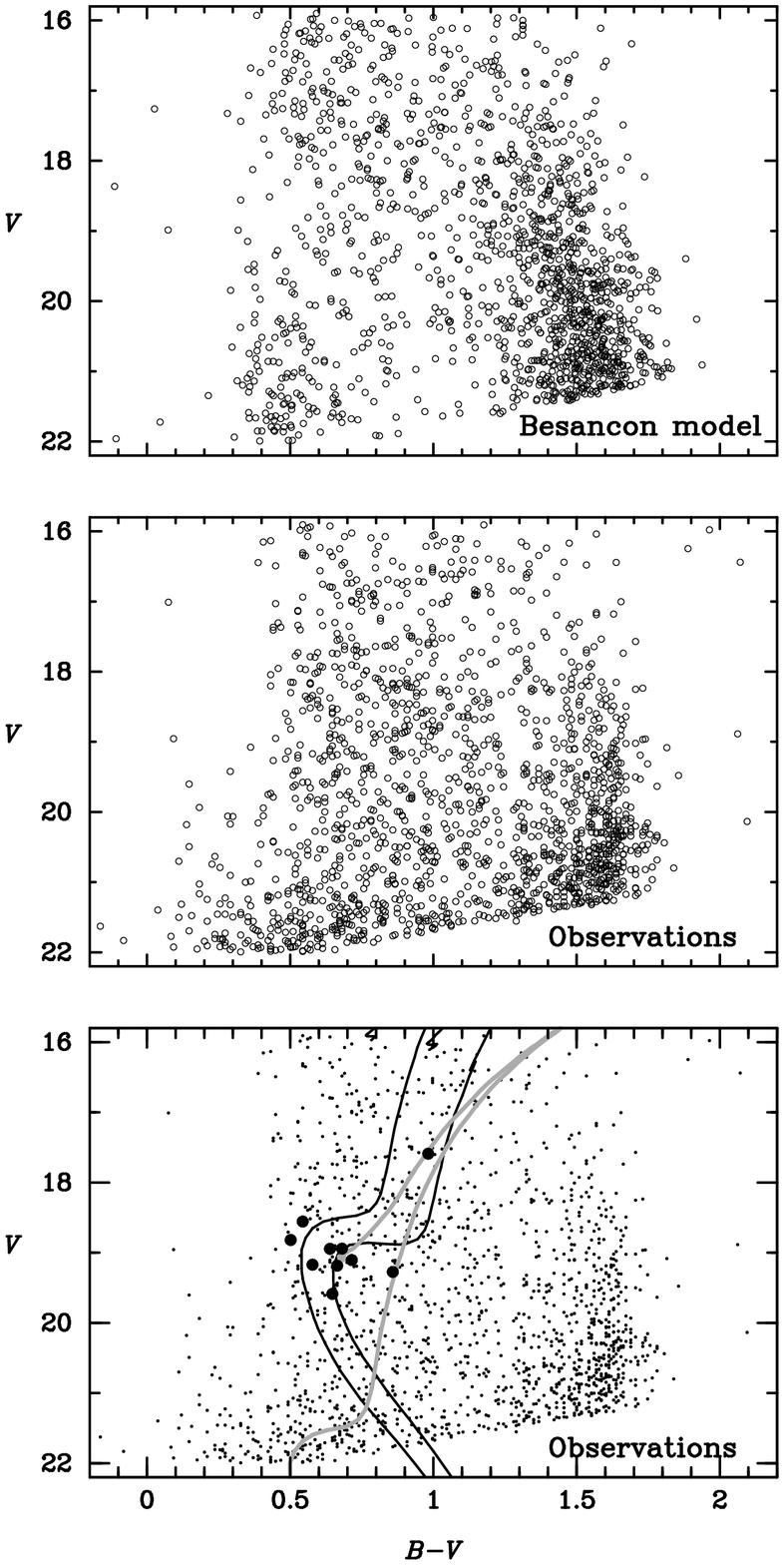}
\caption{CMDs for the Besan\c{c}on model (top) and for our observations (middle and bottom panels).
In addition to the observations, the bottom panel includes two 10 Gyr isochrones at $d_{\odot} = 9$ kpc and [Fe/H] = -1.3 and -0.7 representing Mon,
and one at $d_{\odot} = 39$ kpc and [Fe/H] = -1.7 representing Sgr. The dark symbols in the bottom panel show the Sgr
candidate members according to RVs.}
\end{center}
\end{figure}

%SRM: For Figure 3, can we make smoothed versions of the model and observed CMDs (for the model,
% by running Besancon model many times, for data, Gaussian smoothing the distribution) and then
% subtract one CMD from the other?  Or we could subtract a control field?  Anyway, this would help
% highlight the excesses?

To better illustrate the stellar excess present at blue colors in the observations we show 
$(B-V)$ color distributions for four $V$-magnitude bins in Figure 4. The observations are
represented with a continuous black line, while the model predictions are shown with a grey line.
The magnitude range is specified in each panel. It appears that the Besan\c{c}on model overpredicts
starcounts for stars redder than $B-V = 1.3$ in all magnitude ranges; these stars are red dwarfs 
belonging to the thin disk population. While our purpose here is not to constrain the 
model with our observations, an adjustment of the thin-disk parameters can account for this
discrepancy. 
The stellar excess in the observations when compared to the model predictions is obvious
for colors bluer than $(B-V) < 1.2$, in all three of the faintest magnitude bins.

\begin{figure}[h]
\begin{center}
\includegraphics[scale=1.00]{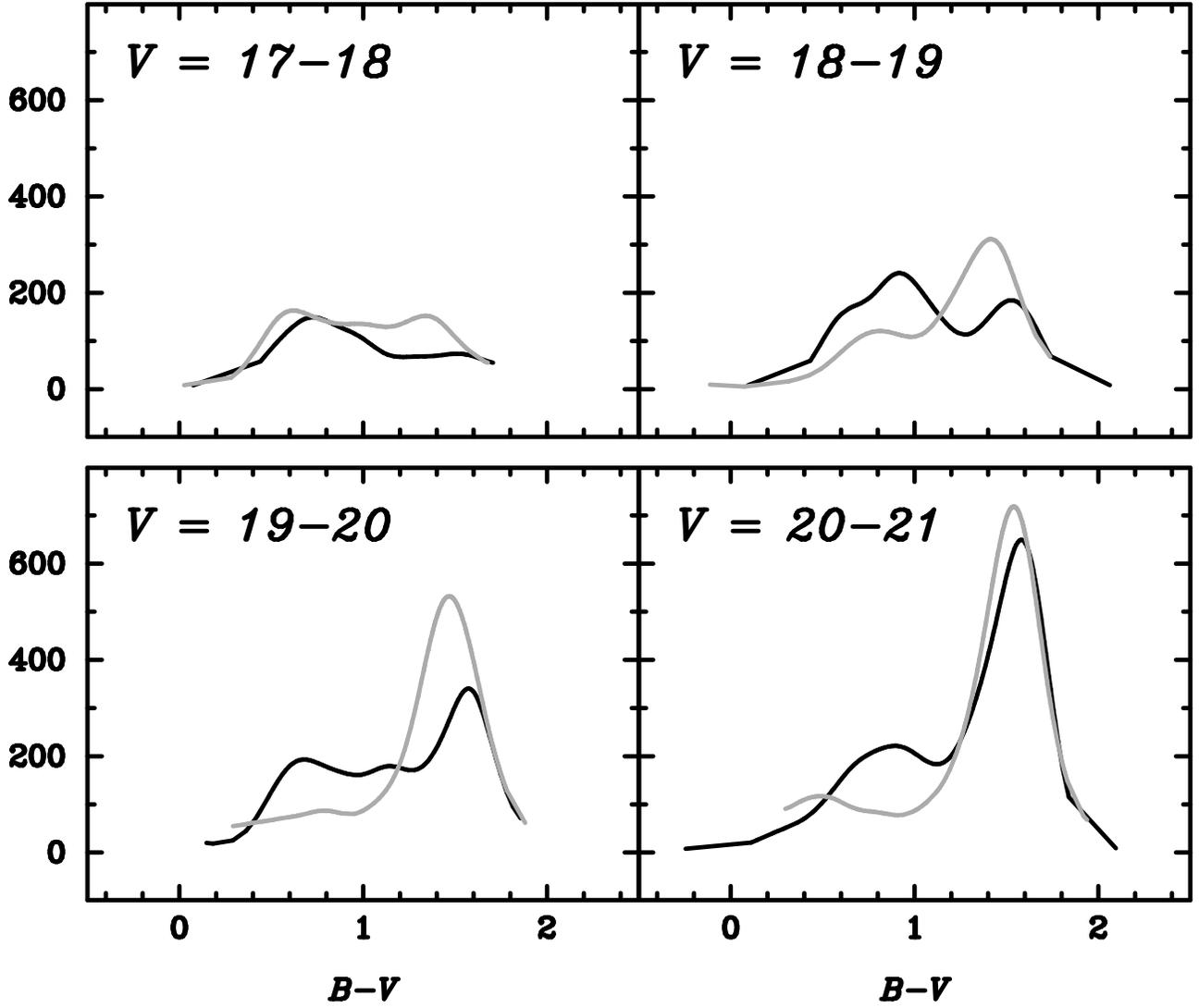}
\caption{Color distributions of the observations (black line) and of the Besan\c{c}on model predictions (grey
line) for four magnitude bins, as specified in each panel.}
\end{center}
\end{figure}

When the D02 paper was written, the 
most well-known disrupting satellite was Sgr. Its orbit projected onto the sky
indicated that Sgr debris can be found in SA 71 (D02), more specifically, debris from the 
southern trailing tail (e.g., Majewski et al. 2003). Additionally,  D02 noted that three carbon stars
from the catalog of Totten and Irwin (1998) that are near SA 71, have highly negative RVs ($\sim -140$ km/s)
that match the values for Sgr tidal debris.
Later, Putman et al. (2004) found an H I stream along the orbit of Sgr, toward the Galactic anticenter
and encompassing SA 71. The highly negative line-of-sight velocity range 
(-380 to -180 km/s, Putman et al. 2004) of the H I stream is 
indicative of a likely membership to the Sgr trailing trail according to the stellar Sgr velocities
 by Majewski et al. (2004). 
Putman et al. (2004) interpreted this H I stream
as a gaseous component of Sgr stripped by ram pressure, as the satellite 
passed through the edge of the Galactic disk. 
In this context, D02 interpreted the stellar excess at 
$(B-V) \sim 1.0$, $V = 18 - 19$, as stellar debris from Sgr, specifically red-clump/horizontal-branch 
stars at $d_{\odot} \sim 30$ kpc. The proper motions too were approximately compatible with 
those predicted for Sgr tidal debris (D02). In the meantime, photometric surveys
revealed another very extended structure: the Monoceros ring. Mon is now best understood as
a tidal feature from a disrupted satellite with a low-inclination, circular orbit
(e.g., P05, Conn et al. 2007 and references therein), although its originating satellite is still a
matter of debate (Conn et al. 2007 and references therein).
SA 71 lies in the general area where Mon debris was
mapped below the Galactic plane (CD06 and references therein), and where the 
P05 model predicts tidal debris. The P05 model predicts Mon debris at a heliocentric
distance $d_{\odot} \sim 9-10 $ kpc, in the region of SA 71. 

Therefore, assuming that in SA 71
we have tidal debris from both Sgr and Mon, we have used the Girardi et al. (2002) isochrones
to qualitatively compare the stellar excess seen in SA 71 to the likely populations of Sgr and Mon.
We have used the 10 Gyr-old isochrones of [Fe/H] = -0.7, -1.3 and -1.7, which are a reasonable approximation for 
the stellar populations in discussion. 
The two most metal rich
isochrones were placed at a distance of 9 kpc, and they were chosen to approximately represent 
populations in Mon (Rocha-Pinto et al. 2003, Ibata et al. 2003,  Yanny et al. 2003).
The most metal poor isochrone was placed at a distance  of
39 kpc, and was chosen to represent Sgr (Newberg et al. 2002), 
although more metal rich stars in Sgr's trailing tail
are also present (Majewski et al. 2003).
The isochrones were shifted to account for the reddening of $E_{B-V} = 0.16$
at the center of our SA 71 field (Schlegel, Finkbeiner \& Davis  1998). The bottom plot of Figure 3
shows our data together with the isochrones. The dark, filled symbols represent
Sgr candidate members according to their RVs (see Section 3.2).
The subgiant, turnoff and possibly main sequence features of the isochrones at 9 kpc
best explain the stellar excess which is redder at $ V = 18-19$ and bluer at $V = 19 - 20$.
Also, the $V$-mag location of the red clump for the isochrone at 39 kpc lies in the 
subgiant/turnoff region of the isochrones representing Mon. As already noted, a more metal rich 
population in Sgr, which is in agreement with the presence of
M giants in the trailing tail (Majewski et al. 2003), would shift the red clump toward redder $B-V$ colors
right in the region were the bright, redder stellar excess was found thus justifying
the D02 interpretation of the excess. 

%SRM: It would seem that Mon would have a significant SPREAD in distance?  Particularly at 9 kpc, 
% a spread of, say, 4 kpc, would be a distance modulus variation of 14.22 to 15.20?  Does this matter?

At fainter magnitudes ($V \ge 21$), there appears to be another prominent
stellar excess at $(B-V) = 0.3 - 0.7$, which can be best interpreted as turnoff stars in Sgr.
While photometric uncertainties are large here (both precision and calibration), and the
stellar/galaxy discrimination is poorer (see Section 2.1), we believe that most objects are stars
at these blue colors. We hope to be able to re-address this issue with deep CCD photometry
in a future contribution.

\subsection{Kinematical Clues to the Stellar Populations and Streams in SA 71}

\subsubsection{Radial Velocities}

%SRM: I moved the former first paragraph here to later on, when it seemed more relevant to bring up?

%At the location of SA 71, the predicted heliocentric radial velocity for Sgr tidal
%%JLC added 'predicted' above
%debris is $\sim -170$ km/s (e.g., Law et al. 2005 models, in a nearly spherical potential), 
%%SRM: I added "in a nearly spherical", assuming you used our q=0.9 model??  But also, don't
%% we have actual Sgr RVs at this position given in Majewski et al. 2004, or Law et al. 2005?
% LATER: Jeff looked into this comparison and it is nice.  We should put a version of his 
% figure in the paper...
%
%and it is
%this velocity component that best separates from the Galactic field population.

In Figure 5, we show the distribution of RVs for the Besan\c{c}on model and for our 
data, as a function of color (left panels) and magnitude (middle panels). Normalized histograms of the 
RV distributions are also plotted in the middle panels of Fig. 5, to show the 
shape of the distributions.
Along with the RV distributions, we show the VPD (right panels) for
each sample. The Besan\c{c}on data were trimmed in color ($0.4 < B-V < 1.2$) and magnitude
( $18.0 < V < 19.7$ ), to match
the selection of our RV sample (Fig. 1, and eliminating stars brighter than $V = 18$). 
No other selection was applied to the Besan\c{c}on
data, while the observed RV sample was selected in proper-motion space to 
preferentially pick out members in the ``cold'' VPD structure. 
%SRM: I added "VPD" for clarity above.
Within the CMD selection $0.4 < B-V < 1.2$ and $18.0 < V < 19.7$  
our observations contain 229 stars 
of which 158 have RVs (see also Fig. 1). For the same CMD selection, the
Besan\c{c}on model has 130 stars.

%SRM: How many stars are in the model in these ranges?  Presumably the excess counts are significant?
% (See my comment elsewhere about statistically subtracting the model and SA71 CMDs?)

\begin{figure}[h]
\includegraphics[scale=0.65,angle=-90]{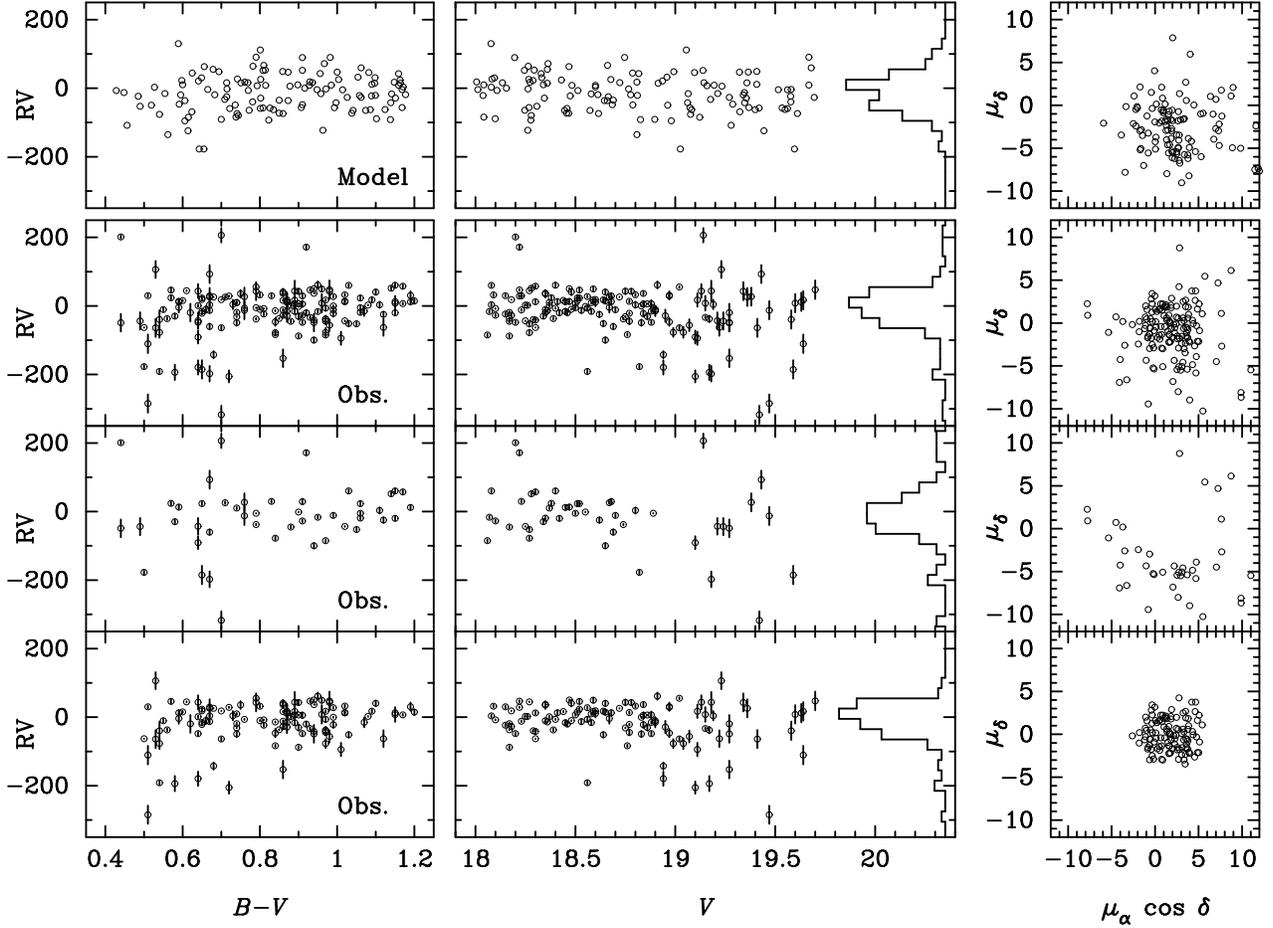}
\caption{RVs as a function of color (left) and magnitude (middle) for the Besan\c{c}on model
(top row) and for various samples of our observations (second to fourth rows). The right panels show 
the proper-motion distributions of the respective samples. The second row shows all of the RV observations,
the third, those outside the proper-motion clump, and the fourth, those inside the proper-motion
clump.}
\end{figure}

The first row of panels in Fig. 5 shows 
the Besan\c{c}on distributions, where errors are convolved into the model, but not represented 
with error bars, for clarity. 
The bottom
three rows of panels show our observed
distributions separated into three groups: all stars with RVs
(second row), stars with RVs and proper motions outside the 
clump in the VPD (third row), and those stars with proper motions
inside the VPD clump (fourth row).  The proper-motion clump is
selected within a $\sim 2\sigma$ radius of the proper-motion centroid.

The first thing that stands out in Figure 5,  is the group of 
stars at RV $\sim -180$ km/s, which also have blue colors and faint magnitudes. 
At the location of SA 71, the predicted heliocentric radial velocity for Sgr tidal
debris is $\sim -170$ km/s (e.g., Law et al. 2005 models, in a spherical potential).
Majewski et al. (2004, their Fig. 2) find similar RV values for M giants in a similar 
part of the Sgr trailing tail.
It is this velocity component that best separates Sgr members from the Galactic field population.
In the 
Besan\c{c}on model, there are two such stars, while in our data there are nine.
The tenth Sgr RV candidate is at $V= 17.6$ (Fig. 3).
If we choose only stars within the tight proper-motion clump, there are six stars with highly
negative RVs that match the velocity predicted for Sgr debris. 
Therefore, in the magnitude range explored here,
a small number of stars in our tight proper-motion clump most likely belong to Sgr.

The majority of the stars in the proper-motion clump, however, are not Sgr members, according to the RVs.
Moreover, they show a smaller RV dispersion than that of the Galactic field (rows 1 and 4 in Fig. 5)
between $V = 18$ and 19. For magnitudes fainter than $V = 19$, uncertainties in RVs increase substantially
(see Fig. 2), and thus we can not characterize well the intrinsic RV dispersion at these magnitudes.
Also, the shape of the RV distribution of the observed data has a sharp
cutoff at RV$\sim 60$ km/s, while that of the Besan\c{c}on model has a gradual decline. The observed 
RV distribution of stars with proper motions outside the VPD clump (Fig. 5, row 3) is more in line 
with the RV distribution of the Besan\c{c}on model. 
The RV dispersion of the excess stars within the proper-motion 
clump is rather difficult to determine accurately, given the overlap with Galactic field
stars in both RV and proper-motion space. Nevertheless, we can estimate it giving 
upper and lower limits as follows. First, from the stars within the proper-motion clump (Fig. 5, bottom row),
we eliminate all stars fainter than $V = 19$, where RV errors increase substantially (Fig. 2), and with 
RV size larger than $100$ km/s. The average for the 81 stars thus selected, is $-0.9 \pm 3.7$ km/s, with 
%JLC By 'RV size', do you mean the absolute magnitude of the RV?  To exclude Sgr *s as well?
$\sigma_{RV} = 33$ km/s. If we consider only stars within $V = 18.3$ to 18.6, where the smallest
scatter is apparent, then we obtain an average $9.9 \pm 4.5$ km/s and a dispersion $\sigma_{RV} = 23$ km/s,
for 27 stars.
Considering that the RV errors are of the order of 10 km/s (Fig. 2), we obtain an intrinsic
RV dispersion between 21 and 31 km/s, which is considerably less than that of the Galactic field
at similar colors and magnitudes, as indicated by the Besan\c{c}on model ($\sim 50$ km/s, Fig. 5, top-middle
panel).
%JLC Need a reference, or some other justification for this 50 km/s Galactic value.
%SRM: Again, would be nice if we had real data to compare to for this -- so we are less model-dependent.
The RVs of the handful of stars observed 
at $V \sim 17.5$ and shown in Fig. 2, also indicate a large ($\sim 40$ km/s) RV dispersion,
more typical of the Galactic field. 

We therefore conclude that the excess blue population between $V = 18$ and 19
is a kinematically cold population compared to that of the
field, but while Sgr debris is present in SA71, it does not dominate the sample.
%it is  present --- Sgr debris does not dominate this population.
%field, and --- while present --- it is not dominated by Sgr debris.
%JLC Changed above sentence.
%SRM: And I changed it again!
%SRM: I think we need a more helpful transition to the possibility of Mon here, so I added the next sentence:
The other obvious possibility is that this concentration in the VPD is a result of the presence of Monoceros
stars.  We note that
radial-velocity studies within Mon overdensities
show RV dispersions that vary between 17 km/s and 26 km/s, with most values
around 22 km/s (Yanny et al. 2003, Crane et al. 2003, Martin et al. 2006); thus our measurements
fit within these determinations.

\subsubsection{Proper Motions}

%SRM: See comment above about recommended section heading changes?

%SRM: Again, some kind of transition or introductory statement would be helpful here?
We proceed now to investigate the proper-motion distributions of all stars ---i.e., not only those with RVs ---
within the general CMD area where the stellar excess was found (Section 3.1).   
Since the RVs indicate that the excess stars are NOT predominantly Sgr stars but rather likely Mon members (Section 3.2),
we select candidate Mon members from the CMD guided by the two isochrones located at 9 kpc
(Fig. 3). First, we select stars within a rectangular area defined 
between $18 \le V \le 21$ and $0.2 \le B-V \le 1.2$. There are 428 stars within this sample.
Second, we use a more restrictive selection 
defined along the two isochrones and allowing for small metallicity and distance variations beyond 
those strictly imposed by the isochrones. This sample is also restricted within the generic magnitude range
$V =18$ to 21, and it includes 213 stars. The CMD selection
is illustrated in the top-left panel of Figure 6.
The open circles show the box cut,
while the filled circles, that along the isochrones. Also, only stars with proper motions 
based on measurements of six or more plates are considered.

\begin{figure}[h]
\includegraphics[scale=0.85]{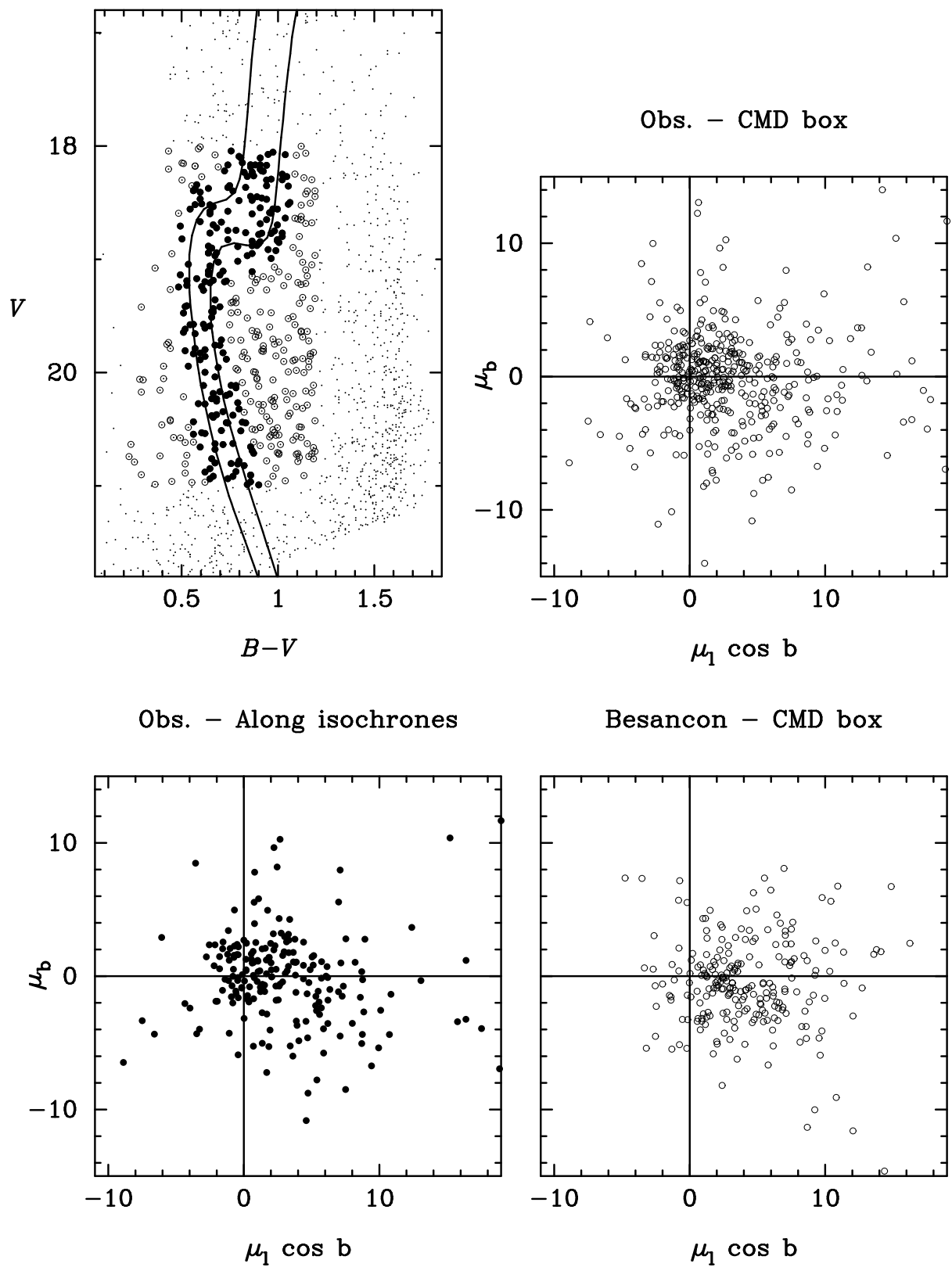}
\caption{The CMD selection (top-left) of candidate Mon members: a box selection (open circles) and 
an isochrone-guided selection (filled circles). The proper-motion distributions of the
CMD-box selected members (top-right) and isochrone-selected members (bottom-left). The Besan\c{c}on model
distribution for the same CMD-box selection is show in the bottom-right panel.}
\end{figure}

%SRM: In Figure 5 it might be a little misleading to the eye that the symbols are different in the different VPDs,
% but I understand why you are doing it.  I just find solid points "look" different (more dense).  BUt I'm not
% recommending a change.  But it might also be useful to show the "non-Mon sample" to compare to the Besancon model??
% Isn't the point that Besancon is meant to show "the field" and you are trying to show the difference with the Mon
% sample?  Useful to show that the "observed field stars" match the Besancon model?   

We show the proper-motion diagrams of these Mon-selected samples in Fig. 6.
These are absolute proper motions, along Galactic coordinates.  The top-right panel
shows the box-selected sample; the bottom-left, that of the sample along isochrones.
The bottom-right panel of Fig. 6 shows the proper-motion distribution given by the 
Besan\c{c}on model within the same CMD box cut applied for the Mon sample selection.
The model predicts only 241 stars in the same CMD region where the observations have 428 stars. We have run the 
Besan\c{c}on model several times; different realizations provide similar numbers within a 
$1\sigma$ uncertainty given by the Poisson statistics.
%
%SRM: I'm a bit surprise by this, in various ways.  FIrst, are you sure this is real in the
% Besancon model?  It looks like a statistical fluctuation in the model, whereas it definitely
% looks real in the data.  Can you run the model multiple times to see if this is just a statistical
% fluctuation?  In any case, one ought to run multiple random seeds to the Besancon model to see
% if things are real or statistical fluctuations.
% But, if it is NOT a fluctuation in the model, then the next obvious question is what the
% heck causes it??  I wasn't aware that the model had any asymmetrical components included -- though
% maybe it has the warp included?  If it *IS* the warp, then this is certainly important, and perhaps
% undercuts our desire to make this Mon debris?  But anyways,
% the the model shows this feature begs the question, why/what is it?  (BTW: It doesn't quite look the same
% as the excess in the data, which makes me wonder if it is a model fluctuation? 
% JLC: Some other concerns about the models: (1) Are we consistent with the model being dereddened,
% but the data not being dereddened?  The E(B-V) = 0.19 so A_V ~ 0.6 in SA71; this is significant?
% (2) As SRM mentioned, running many instances of the model for statistical purposes is important, 
% rather than simply comparing to one run.  This may be especially important when considering
% the elongation in the VPD.   (3) I think the Besancon model includes disk warp/flare components,
% but a detailed reading would be necessary to see how this affects us near SA71.
%
Besides a low-density population with a high proper-motion dispersion, the Besan\c{c}on Galactic-field 
distribution shows an elongated shape, oriented at $\sim 45\arcdeg$ with
respect to the $\mu_l~cos~b$ axis. This proper-motion feature can also be distinguished
in the observed distributions, in both Mon-selected samples. Along with this feature,
there is a significant population clumped around a proper-motion value of
$(\mu_l~cos~b, \mu_b)  \sim (0,0)$ mas/yr,
that is not present in the model. It is this population that we wish to separate from
the Galactic field, and determine its mean absolute proper motion. 

%SRM: I put in a paragraph break here -- the paragraph is pretty long otherwise.
We have tried various procedures
to separate the field distribution from that of the excess population. 
First, we have fitted the observed marginal distributions with the sum of two Gaussians,
one  representing the field, the other the excess population. This is a 
well-documented technique used when separating clusters from the field in proper-motion space
(e.g. Girard et al. 1989). The observed marginal distributions were first 
constructed along $(\mu_l~cos~b, \mu_b)$,
and then along a $45\arcdeg$-rotated system so that the major axis of the field distribution is
along the $x$ proper-motion axis (see Fig. 6). We have worked with both the CMD-box selected sample
%JLC Is 'see Fig. 6' necessary above?  Fig. 6 doesn't actually show
%the rotated system you're talking about here.
and the one selected along the isochrones. Second, we have done a 2-D subtraction 
of the field distribution from the observed distribution, by using
the Besan\c{c}on model to represent the field. We have thus constructed density maps
in the proper-motion space for the CMD box-selected sample for the observed and the Besan\c{c}on
distributions. We subtract the Besan\c{c}on density map from the observed one, and
determine the centroid of the subtracted density map. Two estimates are provided
using this method: one uses the Besan\c{c}on proper-motion distribution as is, assuming
it represents the true Galactic field. The other estimate
uses the Besan\c{c}on proper-motion distribution shifted in proper-motion space so that the
elongated feature of the field in the Besan\c{c}on model is approximately aligned 
with the same feature in the observed space (see Fig. 6). The adopted shift is
determined by eye, and it is $(\mu_l~cos~b, \mu_b) = (0., 1.)$ mas/yr. Internal errors of all
of these estimates are of the order of a few tenth of mas/yr, and much smaller than the
differences between various estimates. 

In Table 2 we list the absolute proper motion
of the excess population obtained from the various estimates described above. Clearly,
the dominant uncertainty in this number is due to the difficulty in 
separating the field population from the excess population.  By using various 
samples and methods, we aim to quantify this uncertainty. 
To this end, we will take the average of the estimates in Tab. 2 as our final
determination of the absolute proper motion of the excess population and the scatter 
(1 standard deviation)
of the various estimates as the uncertainty of the determination. To this 
uncertainty we have to add (in quadrature) the error of the proper-motion
zero point given by the galaxies (Section 2.1).  We thus obtain $(\mu_l~cos~b, \mu_b)
 = (0.72\pm0.72, 0.67\pm0.53)$ mas/yr for the final absolute proper motion estimate of the blue stellar excess in SA 71.

We have inspected a number of other realizations of the Besan\c{c}on model that used the same input
as the realization discussed here. The VPDs of all realizations are statistically consistent, although
the elongated feature discussed above may not distinctly appear in all realizations. This however
does not affect our result within its formal error: for instance, the estimation based only on
the four measurements that do not include the Besan\c{c}on model (Tab. 2) give
 $(\mu_l~cos~b, \mu_b) = (1.06\pm0.64, 0.89\pm0.43)$ mas/yr.

%JLC Should add the final (averaged) value to Table 2.  Otherwise,
%it's hard to find since it's buried in the text.

\begin{table}[htb]
%\begin{center}
\caption{Proper-Motion Estimates} 
\begin{tabular}{lcc}
\tableline \\
\multicolumn{1}{l}{Sample/Method} & \multicolumn{1}{c}{$\mu_l~cos~b$} & \multicolumn{1}{c}{$\mu_b$} \\
\tableline \\
CMD box: $0\arcdeg$  & 0.39  & 0.34 \\
CMD box: $45\arcdeg$ & 1.59  & 0.99 \\ \\
Isochrone: $0\arcdeg$ & 0.68 & 1.27 \\
Isochrone: $45\arcdeg$ & 1.57 & 0.95 \\ \\
2D map - no shift & -0.01 & 0.55 \\
2D map - shift    & 0.12  & -0.11 \\ \\
Average & 0.72    &  0.67 \\
\tableline
\end{tabular}
%\end{center}
\end{table}

\section{Comparison with the Monoceros Tidal Stream Model of Pe\~{n}arrubia et al.}

We compare our data with the disruption model of P05 that is constrained by the
spatial distribution, distance estimates and radial velocities in overdensities
mapped above and below the Galactic plane and between $l = 110\arcdeg - 240\arcdeg$.
We note that the model is not constrained by any data in the neighborhood of
the CMa overdensity (see Fig. 2 in P05). The model that best explained the data set
used by P05 is that of a satellite on a prograde, low-inclination, rather circular orbit.

In Figure 7 we show the spatial distribution in Galactic coordinates of the
tidal stream predicted by P05 (grey points).  The location of our SA 71 field is indicated with a red
symbol. We also show (green line) the orbit of Sgr
based on derived proper motions (Dinescu et al. 2005) 
 integrated back in time for $\sim 2$ Gyr (some three radial periods).
Sgr's orbit precession is from lower $l$ values to higher $l$ values, i.e., moving back in time
the orbit is closer to SA 71. 
From the model, we select a set of points that best represent
SA 71's location in the sky; these are shown with black symbols in Fig. 7.
Our kinematical results for the candidate Mon population are shown in Figure 8, 
together with the model predictions. 
The left panels show the data as a function of $l$, while the right panels show the 
$b$ dependence. Top and middle panels show the proper motions in Galactic longitude and latitude
respectively. The bottom panels show the heliocentric radial velocities.
The observed values are shown with a red symbol, and the dark symbols show the selected 
SA 71-representative sample from the P05 model. For the RV, we indicate the two estimates
from Section 3.2.1. The observations are in good overall agreement with the model predictions.
The mean radial velocity of the model as given by the representative sample, is 
different  from our determination by $\sim 60$ km/s; however our RV measurement is within
the range of predicted values in that region of the sky. Given that the P05 model 
was constrained rather approximately by the data available then, the agreement with 
our observations in all kinematic components is remarkable. Using the 
representative sample from the model, we obtain a RV dispersion of $22$ km/s as given by
probability plots trimmed at $10\%$ for outliers. This value is compatible with our
measurement of the RV dispersion of candidate Mon members (Section 3.2.1). 

\begin{figure}[h]
\includegraphics[scale=0.85]{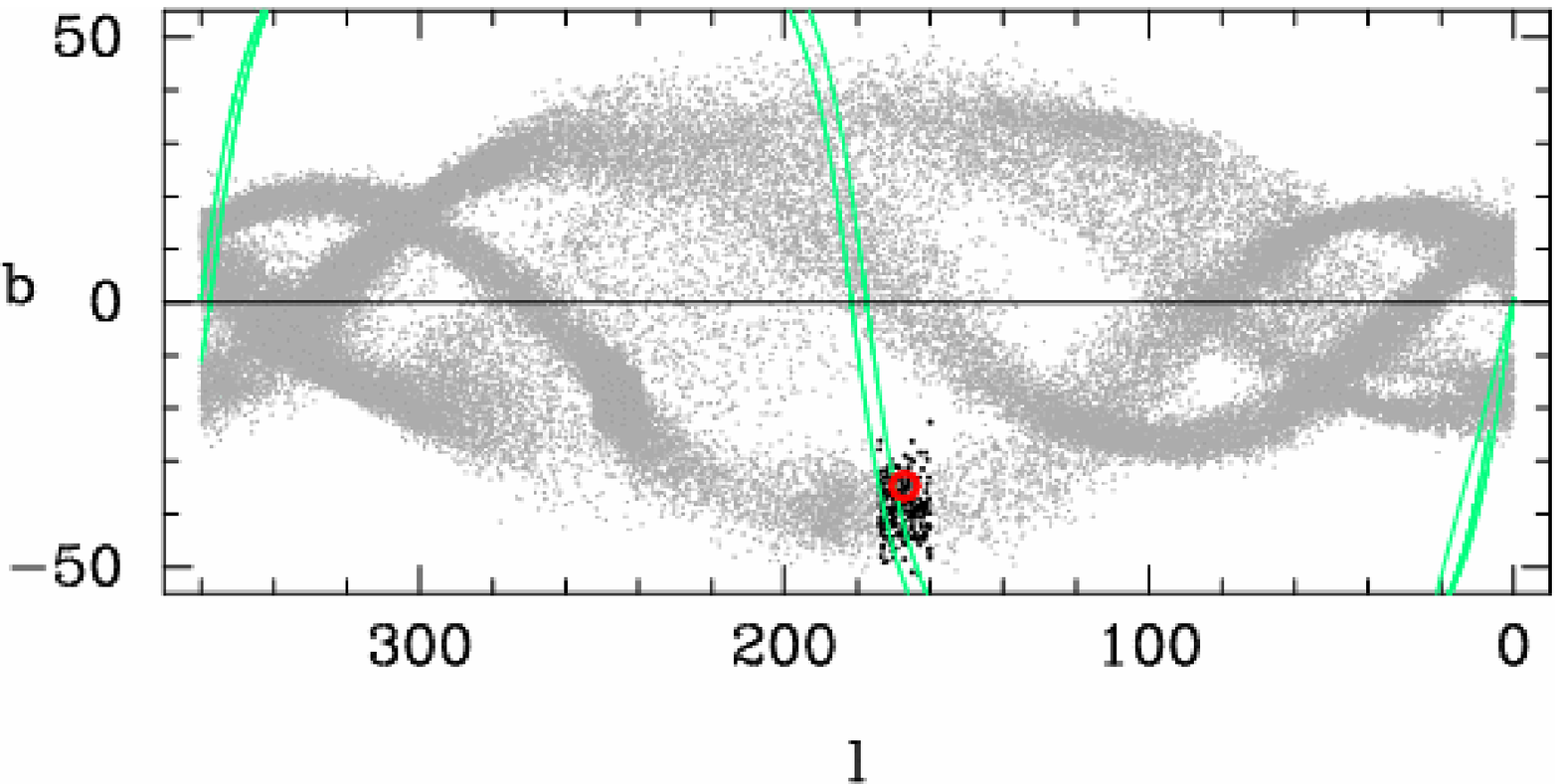}
\caption{The spatial distribution of Mon tidal debris from the model of P05 (grey points).
In green we show the orbit of Sgr, and in red the location of SA 71. The dark points show
the SA71-representative sample selected from the P05 model.}
\end{figure}

\begin{figure}[h]
\includegraphics[scale=0.85]{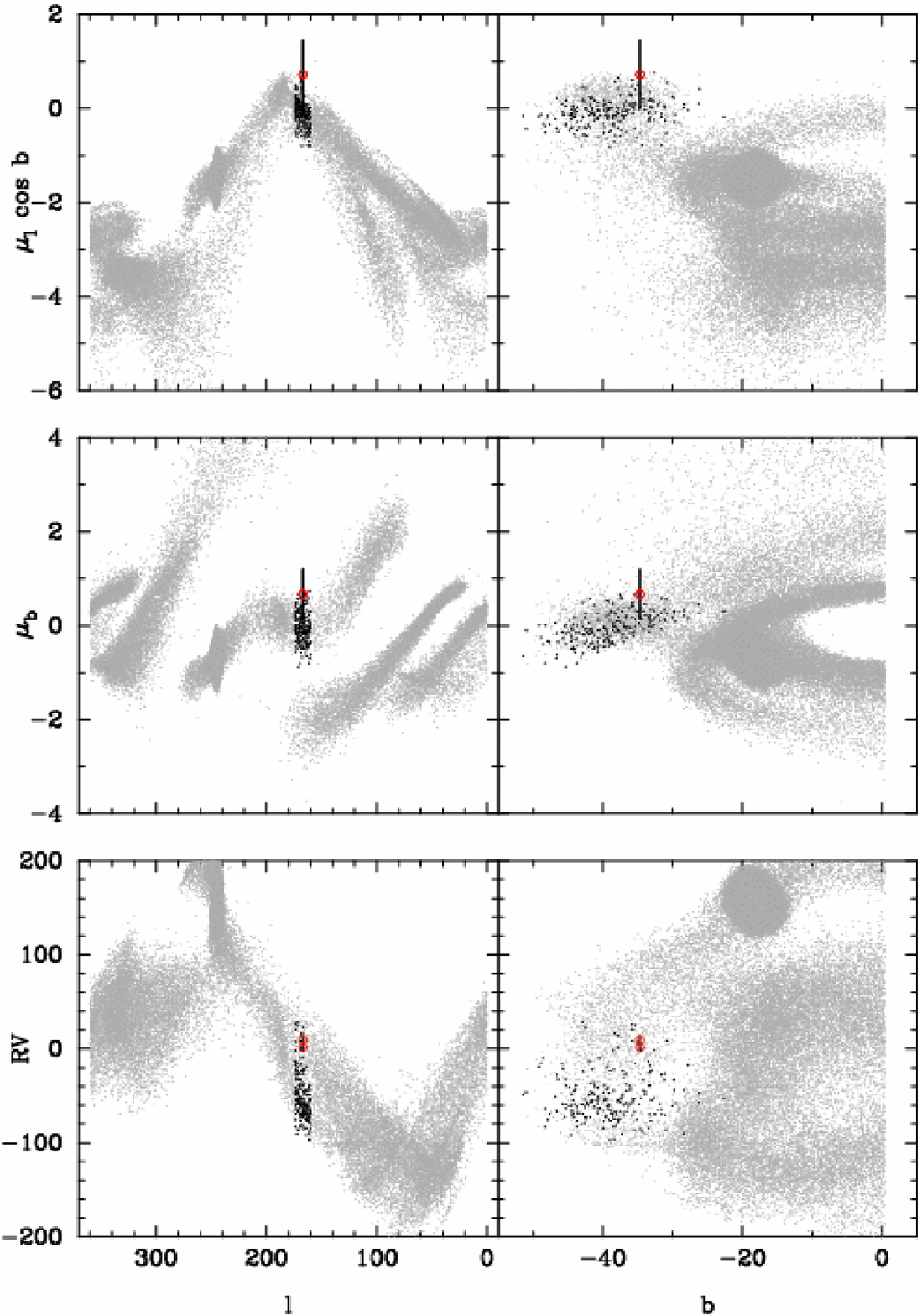}
\caption{Comparison of the P05 kinematical predictions with the observations in SA 71.
The left panels show the dependence with Galactic longitude, while the right panels that with Galactic latitude;
we restrict the comparison to values of $b \le 0\arcdeg$.
The grey points show the model predictions and the dark points those of the SA71-representative sample.
In red we show our determinations in SA 71. The top rows show the proper motion along 
longitude, the  middle, those along latitude, and the bottom the RVs.}
\end{figure}

Another feature of 
the model predictions is the shape of the RV distribution. Within a small region of a 
tidal stream of the model,
the RV distributions show sharp edges on one side of the distribution, and this appears to be
a generic feature in tidal streams.
Our representative sample exemplifies this in 
Fig. 8 - bottom-right panel. This sharp cutoff is also clearly seen in our RV data 
(Fig. 5, middle panels, second and fourth rows). Although
the sharp edge in the observations is on the opposite side of the velocity distribution than in the  
model distribution, we are not concerned here with the exact match of the model, but 
rather with reproducing  "generic" features of the model.

Assuming a distance of $9$ kpc, and adopting the proper motion and radial velocity 
determined here (the low RV estimate), we obtain an orbit with an eccentricity of $0.09 \pm 0.04$ and
an inclination of $20\arcdeg \pm 2\arcdeg$. 
The orbit integration was done in the potential 
from Johnston et al. (1995), and the errors are determined from the formal errors in the observables,
with an assumed $10 \%$ error in the distance (Dinescu et al. 1999).
The higher RV estimate gives an orbit with similar parameters within errors.
The orbit determined here is entirely consistent with that from P05.

\section{Other Considerations: The Galactic Warp or More Nearby Overdensities}

The Milky Way warp has been recently re-mapped from the H I distribution by 
Levine et al. (2006). This work as well as previous others have shown that in the
third quadrant, the disk warps below the Galactic plane, with the line of nodes close
to the direction to the Anticenter.
We also note that the Besan\c{c}on model includes the Galactic warp and the flare as parametrized with 
constraints from the near-infrared DENIS survey (Derri\`{e}re \& Robin 2001).

Nevertheless, we investigate here the
possibility that the stellar excess in SA 71 ($(l,b) = (167.1\arcdeg, -34.7\arcdeg)$, second quadrant)
 may be part of the warp, given that the H I distribution
at longitudes very close to the Anticenter is not well known and that there is an indication that
the disk dips under the Galactic plane close to our region of interest (Fig. 6 in Levine et al. 2006).
If we assume that the distance to our stellar excess is roughly 9 kpc as indicated by the isochrones
(Section 3.1), then this material is at a distance of $\sim 5$ kpc below the Galactic plane.
As indicated by Levine et al. (2006, their Fig. 6), 
the displacement of the H I gas disk below the Galactic plane in the region close 
to SA 71's longitude is at most 1.3 kpc. Therefore the stellar excess in SA 71
 can not be interpreted as material from
the warp, if the assumed distance is 9 kpc. The remaining possibility to investigate is whether
the stellar excess is located more nearby than 9 kpc.

If we aim to account for the stellar excess in the brightest magnitude range,
namely $V = 18$ to 19 (Section 3.1) and that peaks at a color $(B-V) \sim 1.0$,  we can envision
the main sequence of a nearby stellar overdensity that can pass through this region. For this 
exercise, we can use our most metal rich isochrone ([Fe/H] = -0.7) and shift it in magnitude
until the main sequence passes through the color range of the stellar excess. This shift amounts
to $\sim 3$ mags which corresponds to a population located at 2.3 kpc from the Sun, 
and 1.3 kpc below the Galactic plane. 
A more metal rich
population --- that better represents the disk --- would have been shifted less, and therefore the
distance to it would have been larger than 2.3 kpc. The opposite is true for a more metal poor population
than [Fe/H] = -0.7.
Therefore, we can achieve a reasonable displacement below the Galactic plane of about 1.3 kpc or less, if we assume 
the overdensity to be relatively metal poor, i.e., [Fe/H]$\le -0.7$, compared to stars in the disk.
We note however, that at more nearby distances to the Sun, the displacement of the gaseous disk from the
Galactic plane is of the order of a few hundred parsecs (Levine et al. 2006).

The difficulty in considering a more nearby population to explain the stellar excess is not
necessarily the metallicity of the population, but the fact that the color of the 
excess stars is bluer at fainter magnitudes than at brighter magnitudes (Section 3.1, our Fig. 4 and D02). 
This is opposite from what the main sequence 
of a nearby population would imply: at fainter magnitudes we should see an excess of stars at redder colors.

For these reasons, we find it unlikely that the stellar excess is part of the Galactic warp, or a 
stellar overdensity within  2-3 kpc of the Sun.

\section{Conclusions}

Our main findings in area SA 71 that include our previous work (D02) and the current study
are the following. \\
1) We have found a stellar excess of blue stars that is best explained by the 
existence of an old and rather metal-poor population ( $-1.3 \le$ [Fe/H] $ \le -0.7$) located 
at $\sim 9$ kpc. Another possible blue stellar excess can be found in the same 
region, but at larger distances, i.e., 39 kpc.  \\
2) The proper-motion data for the ``bright'' ($V = 18-21$) stellar excess shows a tight clump 
(smeared only by proper-motion uncertainties) that is indicative of a kinematically cold structure.
For the ``faint'' ($V > 21$) stellar excess, the larger proper-motion errors, as well as the poorer
star/galaxy separation do not allow us to draw a definite conclusion. \\
3) The RVs of the stars in the ``bright'' excess indicate the existence of 
a population with a velocity dispersion
between 21 and 31 km/s, which is much lower than that predicted for the Galactic field (50 km/s).
The RV distribution of this population shows a sharp cutoff on one side, unlike the gradual 
decline of the Galactic field distribution, and its mean RV is between -1 and 10 km/s.
 In this stellar excess, we also find ten 
stars with RVs that indicate their membership to the Sgr tidal stream. The Galactic
field should have at most two stars at these extreme RVs (-180 km/s). The magnitudes and colors of the
Sgr stream candidate members are consistent with a red clump/horizontal-branch population 
at some 39 kpc, thus reinforcing the hypothesis that the ``faint'' stellar excess is
indeed due to turnoff/main sequence stars in the  Sgr stream. \\
4) After separating from the field population, the mean proper motion and RV 
of the stars in the ``bright'' excess are
compatible with the model predictions for the Monoceros stream by P05. 
Under the assumption that the debris is located at 9 kpc --- another P05 prediction for SA 71 
that accounts for the magnitudes and colors of the ``bright'' excess stars --- we
determine a prograde orbit of low-inclination and rather circular shape that is consistent
with the orbit determined by P05. \\
5) The sharp edge of the observed RV distribution is similar to that predicted by the P05 model
in portions of the stream, and unlike that of the Galactic field. The observed RV dispersion is
within the values determined by other studies in different regions of Mon. \\
6) The uncertainty in the determination of the proper motion and RV for the candidate Mon members is due to the
significant overlap with the field population in all three velocity components. \\

Financial support from NSF grants AST-0406884
and AST-0407207 for this research is acknowledged.
J.L.C. is supported by the Virginia Space Grant Consortium.

\end{document}